\documentclass[10pt,aps,superscriptaddress,reprint]{revtex4-1}
\usepackage{amsmath}
\usepackage{graphicx}
\usepackage[section]{placeins}
\usepackage{amsmath}
\usepackage{amssymb}
\usepackage{amsfonts}
\usepackage{dsfont}
\newcommand{\ket}[1]{|#1\rangle}             
\newcommand{\op}[1]{\widehat{#1}}         
\begin{document}
\title{Polarization pattern of vector vortex beams generated by $q$-plates with different topological charges}
\author{Filippo Cardano}
\affiliation{Dipartimento di Scienze Fisiche, Universit\`{a} di Napoli ``Federico II'', Complesso di Monte S. Angelo, 80126 Napoli, Italy}
\author{Ebrahim Karimi}
\email{Corresponding author: karimi@na.infn.it}
\affiliation{Dipartimento di Scienze Fisiche, Universit\`{a} di Napoli ``Federico II'', Complesso di Monte S. Angelo, 80126 Napoli, Italy}
\author{Sergei Slussarenko}
\affiliation{Dipartimento di Scienze Fisiche, Universit\`{a} di Napoli ``Federico II'', Complesso di Monte S. Angelo, 80126 Napoli, Italy}
\author{Lorenzo Marrucci}
\affiliation{Dipartimento di Scienze Fisiche, Universit\`{a} di Napoli ``Federico II'', Complesso di Monte S. Angelo, 80126 Napoli, Italy}
\affiliation{CNR-SPIN, Complesso Universitario di Monte S. Angelo, 80126 Napoli, Italy}
\author{Corrado de Lisio}
\affiliation{Dipartimento di Scienze Fisiche, Universit\`{a} di Napoli ``Federico II'', Complesso di Monte S. Angelo, 80126 Napoli, Italy}
\affiliation{CNR-SPIN, Complesso Universitario di Monte S. Angelo, 80126 Napoli, Italy}
\author{Enrico Santamato}
\affiliation{Dipartimento di Scienze Fisiche, Universit\`{a} di Napoli ``Federico II'', Complesso di Monte S. Angelo, 80126 Napoli, Italy}
\affiliation{CNISM-Consorzio Nazionale Interuniversitario per le Scienze Fisiche della Materia, Napoli, Italy}
\begin{abstract}
We describe the polarization topology of the vector beams emerging from a patterned birefringent liquid crystal plate with a topological charge $q$ at its center ($q$-plate). The polarization topological structures for different $q$-plates and different input polarization states have been studied experimentally by measuring the Stokes parameters point-by-point in the beam transverse plane. Furthermore, we used a tuned $q=1/2$-plate to generate cylindrical vector beams with radial or azimuthal polarizations, with the possibility of switching dynamically between these two cases by simply changing the linear polarization of the input beam.\newline
OCIS number: 050.4865, 260.6042, 260.5430, 160.3710.
\end{abstract}
\maketitle
\section{Introduction}
The polarization of light is a consequence of the vectorial nature of the electromagnetic field and is an important property in almost every photonic application, such as imaging, spectroscopy, nonlinear optics, near-field optics, microscopy, particle trapping, micromechanics, etc. Most past research dealt with scalar optical fields, where the polarization was taken uniform in the beam transverse plane. More recently, the so-called vector beams were introduced, where the light polarization in the beam transverse plane is space-variant~\cite{zhan09}. As compared with homogeneously polarized beams, vector beams have unique features. Of particular interest are the singular vector beams where the polarization distribution in the beam transverse plane has a vectorial singularity as a C-point or L-line, where the azimuth angle and orientation of polarization ellipses are undefined, respectively~\cite{nye99,dennis09}. The polarization singular points are often coincident with corresponding singular points in the optical phase, thus creating what are called vector vortex beams. Vector vortex beams are strongly correlated to singular optics, where the optical phase at a zero point of intensity is undetermined~\cite{soskin01} and to light beams carrying definite orbital angular momentum (OAM)~\cite{sonja08}. Among vector vortex beams, radially or azimuthally polarized vector beams have received particular attention for their unique behavior under focusing~\cite{dorn03,wang08,zhan02} and have been proved to be useful for many applications such as particle acceleration~\cite{varin02}, single molecule imaging~\cite{novotny01}, nonlinear optics~\cite{bouhelier03}, Coherent anti-Stokes Raman scattering microscopy~\cite{lu09}, and particle trapping and manipulation~\cite{zhan04}. Because of their cylindrical symmetry, the vector beams with radial and azimuthal polarization are also named cylindrical vector beams~\cite{zhan09}.

The methods to produce vector beams can be divided into active and passive. Active methods are based on the output of novel laser sources with specially designed optical resonators\cite{oron00,kozawa05,kawauchi08}. The passive methods use either interferometric schemes~\cite{wang07}, or mode-forming holographic and birefringent elements~\cite{churin93,bomzon02,dorn03,machavariani08,fadeyeva10,stalder96}. Light polarization is usually thought to be independent of other degrees of freedom of light, but it has been shown recently that photon spin angular momentum due to the polarization can interact with the photon orbital angular momentum when the light propagate in a homogenous~\cite{brasselet09} and an inhomogenous birefringent plate~\cite{marrucci06prl,marrucci11jo}. Such interaction, indeed, gives the possibility to convert the photon spin into orbital angular momentum and \textit{viceversa} in both classical and quantum regimes~\cite{nagali09opex}.
\begin{figure*}[!htbp]
\begin{center}
    \includegraphics[width=18cm,draft=false]{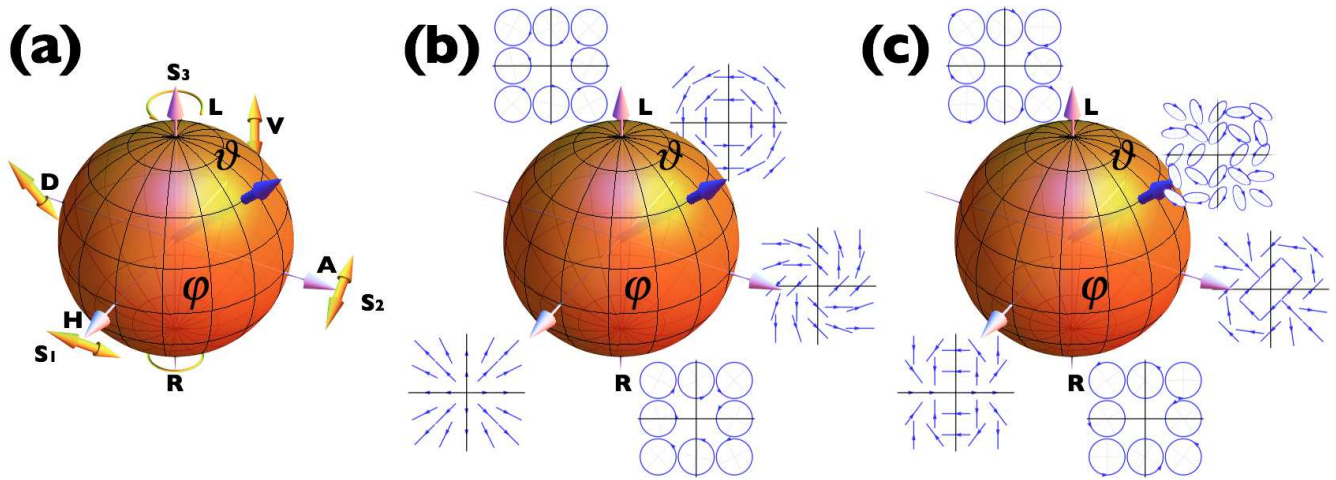}
    \caption{\label{fig:fig1} (Color online) (a) Poincar\'{e} sphere representation for the polarizaiton af a light beam with uniform transverse phase distribution. North (south) pole and equator correspond to left (right)-circular and linear polarizations, respectively. The other points have elliptical polarization, and antipodal points are orthogonal to each other. (b) and (c) show the polarization distribution in the transverse plane for $m=1$ and $m=2$ on the higher-order Poincar\'{e} sphere introduced by Milione et~al.~\cite{milione11prl}, respectively.}
\end{center}
\end{figure*}

In this work, to create optical vector beams we exploit the spin-to-orbital angular momentum coupling in a birefringent liquid crystal plate with a topological charge $q$ at its center, named ``$q$-plate''~\cite{marrucci06prl,marrucci11jo}. As it will be shown later, there is a number of advantages in using $q$-plates, mainly because the polarization pattern impressed in the output beam can be easily changed by changing the polarization of the incident light~\cite{marrucci06apl,karimi10praa}, and $q$-plates can be easily tuned to optimal conversion by external temperature~\cite{karimi09apl} or electric fields~\cite{piccirillo10apl,slussarenko11opex}. Subsequently, the structure and quality of the produced vector field have been analyzed by point-by-point Stokes parameters tomography in the beam transverse plane for different $q$-plates and input polarization states. In particular, we generated and studied in detail the radial and azimuthal polarizations produced by a $q$-plate with fractional topological charge $q=1/2$.

\section{Polarization topology}
Henry Poincar\'e presented a nice pictorial way to describe the light polarization based on the $2\rightarrow1$ homomorphism of SU(2) and SO(3). In this description, any polarization state is represented by a point on the surface of a unit sphere, named ``\textit{Poincar\'e}'' or ``\textit{Bloch}'' sphere. The light polarization state is defined by two independent real variables $(\vartheta,\varphi)$, ranging in the intervals $[0,\pi]$ and $[0,2\pi]$, respectively, which fix the colatitude and azimuth angles over the sphere. An alternative algebraic representation of the light polarization state in terms of the angles $(\vartheta,\varphi)$ is given by
\begin{eqnarray}\label{eq:state}
    \ket{\vartheta,\varphi}=\cos{\left(\frac{\vartheta}{2}\right)}\ket{L}+e^{i\varphi}\,\sin{\left(\frac{\vartheta}{2}\right)}\ket{R},
\end{eqnarray}
where  $\ket{L}$ and $\ket{R}$ stand for the left and right-circular polarizations, respectively. On the Poincar\'{e} sphere, north and south poles correspond to left and right-circular polarization, respectively, while any linear polarization lies on the equator, as shown in Fig.~\ref{fig:fig1}-(a). Special linear polarization states are the $\ket{H}$, $\ket{V}$, $\ket{D}$, $\ket{A}$, which denote horizontal, vertical, diagonal and anti-diagonal polarizations, respectively. In points different from the poles and the equator the polarization is elliptical with left (right)-handed ellipticity in the north (south) hemisphere.

An alternative mathematical description of the light polarization state, which is based on SU(2) representation, was given by George Gabriel Stokes in 1852. In this representation, four parameters $S_i$ ($i=0,\ldots,3$) known as Stokes parameters nowadays, are exploited to describe the polarization state. This representation is useful, since the parameters $S_i$ are simply related to the light intensities $I_p (p=L,R,H,V,D,A)$ measured for different polarizations, according to
\begin{eqnarray}\label{eq:stokes}
	S_0&=&I_L+I_R\cr
	S_1&=&I_H-I_V\cr
	S_2&=&I_A-I_D\cr
	S_3&=&I_L-I_R.
\end{eqnarray}
The Stokes' parameters can be used to describe partial polarization too. In the case of fully polarized light, the reduced Stokes parameters $s_i=S_i/S_0 (i=1,2,3)$ can be used, instead. We may consider the reduced parameters $s_i$ as the Cartesian coordinates on the Poincar\'{e} sphere. The $s_i$ are normalized to $\sum_{i=1}^3s_i^2=1$. The states of linear polarization, lying on the equator of the Poincar\'{e} sphere, have $s_3=0$. The two states $s_3=\pm1$ correspond to the poles and are circularly polarized. In singular optics, these two cases may form two different type of polarization singularities. For the other states, the sign of $s_3$ fixes the polarization helicity; left-handed for $s_3>0$ and right-handed for $s_3<0$.  The practical advantage of using the parameters $s_i$ is that they are dimensionless and depend on the ratio among intensities. Light intensities can be easily measured by several photodetectors and can be replaced by average photon counts in the quantum optics experiments. Thus, Stokes' analysis provide a very useful way to perform the full tomography of the polarization state (\ref{eq:state}) in both classical and quantum regimes.

The passage of light through optical elements may change its polarization state. If the optical element is fully transparent, the incident power is conserved and only the light polarization state is affected. The action of the transparent optical element is then described by a unitary transformation on the polarization state $\ket{\vartheta,\varphi}$ in Eq.~(\ref{eq:state}) and corresponds to a continuous path on the Poincar\'{e} sphere. In most cases, the optical element can be considered so thin that the polarization state is seen to change abruptly from one point $P_1$ to a different point $P_2$ on the sphere. In this case, it can be shown that the path on the sphere is the geodetic connecting $P_1$ to $P_2$~\cite{bhandari97}. Examples of devices producing a sudden change of the light polarization in passing through are half-wave (HW) and quarter-wave (QW) plates. A sequence of HW and QW can be used to move the polarization state on the whole Poincar\'{e} sphere, which corresponds to arbitrary SU(2) transformation applied to the $\ket{\vartheta,\varphi}$ state in Eq.~(\ref{eq:state}). A useful sequence QW-HW-QW-HW (QHQH) to perform arbitrary SU(2) transformation on the light polarization state is presented in Ref~\cite{karimi10praa}.

So far we considered an optical phase that is uniform in the beam transverse plane. Allowing for a nonuniform distribution of the optical phase between different electric field components gives rise to polarization patterns, like azimuthal and radial ones, where special topologies appears in the transverse plane. The topological structure of the polarization distribution, moreover, remains unchanged while the beam propagates. It is worth noting that most singular polarization patterns in the transverse plane can still be described by polar angles $\vartheta,\varphi$ on the Poincar\'{e} sphere~\cite{milione11prl}. The points on the surface of this higher-order Poincar\'{e} sphere represent polarized light states where the optical field changes as $e^{\pm im\phi}$, where $m$ is a positive integer and $\phi=\arctan(y/x)$ is the azimuthal angle in the beam transverse plane. As it is well known, light beams with optical field proportional to $e^{im\phi}$ are vortex beams with topological charge $m$, which carry a definite OAM $\pm m\hbar$ per photon along their propagation axis. Because the beam is polarized, it carries spin angular momentum (SAM) too, so that the photons are in what may be called a spin-orbit state. Among the spin-orbit states, only a few states can be described by the higher-order Poincar\'{e} sphere and, precisely, the states belonging to the spin-orbit SU(2)-subspace spanned by the two base vectors $\{\ket{L}\,e^{im\phi},\ket{R}\,e^{-im\phi}\}$. In this representation the north pole, south pole and equator correspond to the base state $\{\ket{L}\,e^{im\phi}\}$, the base state $\{\ket{R}\,e^{-im\phi}\}$, and linear polarization with rotated topological structure of charge $m$, respectively. Fig.~\ref{fig:fig1}b,c show the polarization distribution for $m=1$ and $m=2$ spin-orbit subspaces, respectively. The intensity profile for all points on the higher-order Poincar\'{e} sphere has the same doughnut shape. The states along the equator are linearly polarized doughnut beams with topological charge $m$, differing in their orientation only.

\section{The $q$-plate: patterned liquid crystal cell for generating and manipulating helical beam}
The $q$-plate is a liquid crystal cell patterned in specific transverse topology, bearing a well-defined integer or semi-integer topological charge at its center~\cite{marrucci11jo,marrucci06prl,marrucci06apl}.

The cell glass windows are treated so to maintain the liquid crystal molecular director parallel to the walls (planar strong anchoring) with non-uniform orientation angle $\alpha=\alpha(\rho,\phi)$ in the cell transverse plane, where $\rho$ and $\phi$ are the polar coordinates. Our $q$-plates have a singular orientation with topological charge $q$, so that $\alpha(\rho,\phi)$ is given by
\begin{eqnarray}\label{eq:qplate}
    \alpha(\rho,\phi) = \alpha(\phi) = q\phi+\alpha_0,
\end{eqnarray}

with integer or semi-integer $q$ and real $\alpha_0$. This pattern was obtained with an azo-dye photo-alignment technique~\cite{slussarenko11opex}. Besides its topological charge $q$, the behavior of the $q$-plate depends on its optical birefringent retardation $\delta$. Unlike other LC based optical cells~\cite{stalder96} used to produce vector vortex beam, the retardation $\delta$ of our $q$-plates can be controlled by temperature control or electric field~\cite{karimi09apl,piccirillo10apl}. A simple argument based on Jones matrix shows that the unitary action $\op{\mbox{U}}$ of the $q$-plate in the state (\ref{eq:state}) is defined by
\begin{equation}\label{eq:q-plate_action}
    \op{\mbox{U}}\left(\begin{array}{c}\ket{L} \\\ket{R}\end{array}\right)=
         \cos\frac{\delta}{2}\left(\begin{array}{c}\ket{L} \\\ket{R}\end{array}\right)+
         \sin\frac{\delta}{2}\left(\begin{array}{c}\ket{R}e^{+2i(q\phi+\alpha_0)} \\\ket{L}e^{-2i(q\phi+\alpha_0)}\end{array}\right).
\end{equation}
The $q$-plate is said to be tuned when its optical retardation is $\delta=\pi$. In this case, the first term of Eq.~(\ref{eq:q-plate_action}) vanishes and the optical field gains a helical wavefront with double of the plate topological charge ($2q$). Moreover, the handedness of helical wavefront depends on the helicity of input circular polarization, positive for left-circular and negative for right-circular polarization.
\begin{figure}[t]
\begin{center}
    \includegraphics[width=8.8cm,draft=false]{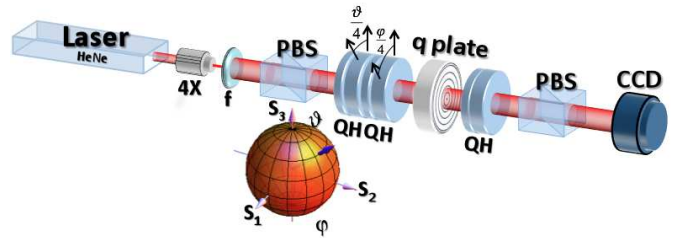}
    \caption{\label{fig:fig2} (Color online) Setup to generate and analyze different polarization topologies generated by a $q$-plate. The polarization state of the input laser beam was prepared by rotating the two half-wave plates in the QHQH set at angles $\vartheta/4$ and $\varphi/4$ to produce a corresponding rotation of $(\vartheta,\varphi)$ on the Poincar\'{e} sphere as indicated in the inset. The waveplates and polarizer beyond the $q$-plate where used to project the beam polarization on the base states (R, L, H, V, A, D) shown in Fig.~\ref{fig:fig1}a. For each state projection, the intensity pattern was recorded by CCD camera and the signals were analyzed pixel-by-pixel to reconstruct the polarization pattern in the beam transverse plane. Legend: 4X~-~microscope objective of 4X used to clean the laser mode, f~-~lens, Q~-~quarter wave plate, H~-~half wave plate, PBS-polarizing beam-splitter.}
\end{center}
\end{figure}

\section{Experiment}
\begin{figure}[!htbp]
\begin{center}
    \includegraphics[width=8.8cm,draft=false]{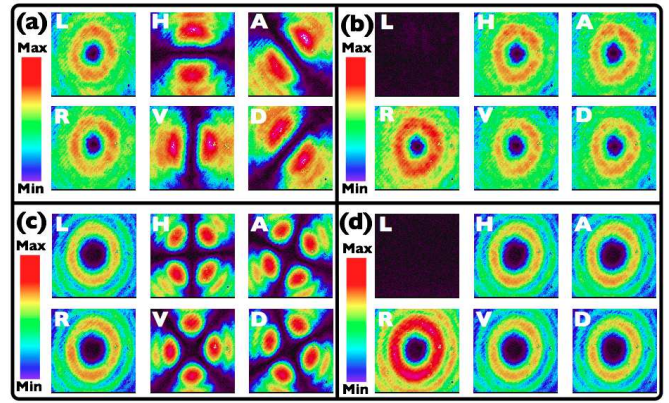}
    \caption{\label{fig:fig3} (Color online) Recorded intensity profiles of the beam emerging from the $q$-plate projected over horizontal (H), vertical (V), anti-diagonal (A), diagonal (D), left-circular (L) and right-circular (R) polarization base states for different $q$-plates and input polarizations. (a) and (b) are for the $q=1/2$-plate, and horizontal-linear (a) and left-circular (b) polarization of the input beam. (c) and (d) are the same for the $q=1$-plate. The color scale bar shows the intensity scale (arbitrary units) in false colors.}
\end{center}
\end{figure}

In our experiment, a TEM$_{00}$ HeNe laser beam ($\lambda=632.8$nm, $10$mW) was spatially cleaned by focusing into a $50\mu m$ pinhole followed by a truncated lens and polarizer, in order to have a uniform intensity and a homogeneous linear polarization. The beam polarization was then manipulated by a sequence of wave plates as in Ref.~\cite{karimi10praa} to reach any point on the Poincar\'{e} sphere. The beam was then sent into an electrically driven $q$-plate, which changed the beam state into an entangled spin-orbit state as given by Eq.~\ref{eq:q-plate_action}. When the voltage on the $q$-plate was set for the optical tuning, the transmitted beam acquired the characteristic doughnut shape with a hole at its center. The output beam was analyzed by point-to-point polarization tomography, by projecting its polarization into the H, V,  A, D, L, R sequence of basis and measuring the corresponding intensity at each pixel of a $120\times120$ resolution CCD camera (Sony AS-638CL). Examples of the recorded intensity profiles are shown in Fig.~\ref{fig:fig3}. A dedicated software was used to reconstruct the polarization distribution on the beam transverse plane. To minimize the error due to small misalignment of the beam when the polarization was changed, the values of the measured Stokes parameters were averaged over a grid of $20\times20$ squares equally distributed over the image area. No other elaboration of the raw data nor best fit with theory was necessary.

We analyzed the beams generated by two different $q$-plates with charges $q=1/2$ and $q=1$ for two different input polarization states. The $q$-plate optical retardation was optimally tuned for $\lambda=632.8$nm by applying an external voltage of a few volt~\cite{slussarenko11opex}. We considered left-circular ($\ket{L}$) and linear-horizontal ($\ket{H}$) polarized input beams. These states, after passing through the $q$-plate are changed into $\ket{R,+2q}$ and $\left(\ket{R,+2q}+\ket{L,-2q}\right)/\sqrt{2}$, respectively. Figure \ref{fig:fig3} shows the output intensity patterns for different polarization selections. Figure~\ref{fig:fig3} (a), (b) show the results of point-by-point polarization tomography of the output from $q=1/2$-plate for left-circular and horizontal-linear input polarizations, respectively. Figure~\ref{fig:fig3} (c), (d) show the results of point-by-point polarization tomography of the output from $q=1$-plate  for left-circular and horizontal-linear polarizations, respectively.

\begin{figure*}[t]
\begin{center}
    \includegraphics[width=17cm,draft=false]{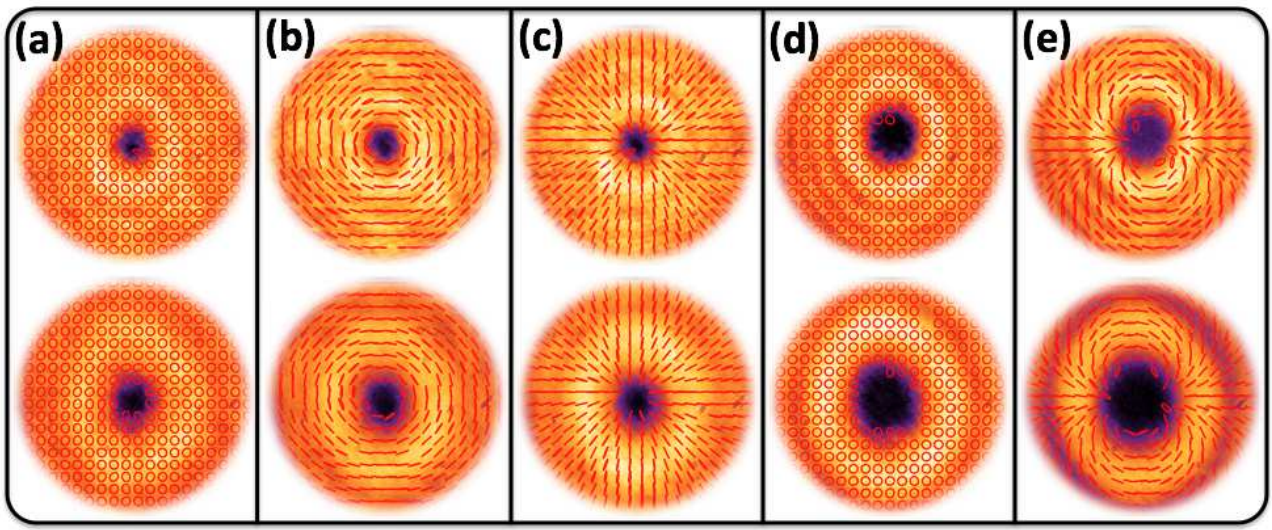}
    \caption{\label{fig:fig4} (Color online) Highest-row panels: the transverse polarization and intensity distribution in the near field at the exit face of the $q$-plate. Lower panels: the polarization and intensity distributions in the far field beyond the $q$-plate. (a), (b), and (c): polarization topological structure generated by the $q=1/2$-plate for left-circular, V-linear, and H-linear input polarizations, respectively. (d)-(e): polarization topological structure generated by the $q=1$-plate for left-circular and H-linear input polarizations, respectively. (a) and (d) have uniform circular polarization distributions.}
\end{center}
\end{figure*}

The case of $q=1/2$-plate is particularly interesting, because for a linear horizontal input polarization, it yields to the spin-orbit state $\left(\ket{R,+1}+\ket{L,-1}\right)/\sqrt{2}$ represented by the $S_1$-axes over equator of higher-order Poincar\'e sphere (Fig.~\ref{fig:fig2} (b)), which corresponds to a radially polarized beam, as shown in Fig.~\ref{fig:fig4} (c). This radial polarization can be changed into the azimuthal polarization (corresponding to the antipodal point on $S_1$-axes of the higher-order Poincar\'{e} sphere) by just switching the input linear polarization from horizontal to vertical, as it is shown in Fig.~\ref{fig:fig4} (b). This provides a very fast and easy way to switch from radial to azimuthal cylindrical vector beam. As previously said, cylindrical vector beams have a number of applications and can be used to generate uncommon beams such as electric and magnetic needle beams, where the optical field is confined below diffraction limits. Such beams have a wide range of applications in optical lithography, data storage, material processing, and optical tweezers~\cite{dorn03,wang08}.

Before concluding, it is worth of mention that vortex vector beams are based on non-separable optical modes, which is itself an interesting concept in the framework of classical optics. At the single photon level, however, the same concept has even more fundamental implications, because it is at the basis of the so-called quantum contextuality~\cite{karimi10prab}.
\section{Conclusion}
We have generated and analyzed a few vector vortex beams created by a patterned liquid crystal cell with topological charge, named $q$-plate. Radial and azimuthal cylindrical beams have been obtained by acting on the polarization of a traditional laser beam sent through a $q=1/2$-plate. In this way, fast switching from the radial to the azimuthal polarization can be easily obtained. Finally, we studied in detail the polarization of a few vector beams generated by different $q$-plates and the polarization distribution patterns have been reconstructed by point-by-point Stokes' tomography over the entire transverse plane.\newline In this paper, however, we have investigated the cases $q = 1$ and $q = 1/2$. Future work will address other cases, and in particular the negative $q$ ones.

\section{Acknowledgment}
We acknowledge the financial support of the FET-Open program within the 7th Framework Programme of the European Commission under Grant No. 255914, Phorbitech.

\end{document}